\documentclass[twocolumn,superscriptaddress,showpacs]{revtex4}
\usepackage{amsmath,amssymb,amsfonts,color,graphicx,bm}

\begin{document}
\title{Random Symmetry Breaking and Freezing in Chaotic Networks }
\author{Y. Peleg}
\affiliation{Department of Physics, Bar-Ilan University, 52900 Ramat-Gan, Israel}
\author {W. Kinzel}
\affiliation{Institute for Theoretical Physics, University of W\"{u}rzburg, Am Hubland, 97074 W\"{u}rzburg, Germany}
\author {I. Kanter}
\affiliation{Department of Physics, Bar-Ilan University, 52900 Ramat-Gan, Israel}

\begin{abstract}
Parameter space of a driven damped oscillator in a double well potential presents either a chaotic trajectory with sign oscillating amplitude or a non-chaotic trajectory with a fixed sign amplitude.
A network of such delay coupled damped oscillators is shown to present chaotic dynamics while the amplitude sign of each damped oscillator is randomly frozen.
This phenomenon of random broken global symmetry of the network simultaneously with random freezing of each degree of freedom is accompanied by the existence of exponentially many randomly frozen chaotic attractors with the size of the network.
Results are exemplified by a network of modified Duffing oscillators with infinite range pseudo-inverse delayed interactions.
\end{abstract}

\maketitle

Chaos and spontaneous symmetry breaking seem to exclude themselves.
A chaotic system explores its complete available phase space \cite{Schuster}, hence multiple attractors occur for very special systems, only \cite{Yanchuk}.
In particular, when a nonlinear system has a plus/minus symmetry, any initial state with a positive amplitude has a twin with a corresponding negative amplitude.
Any chaotic trajectory will visit positive and negative amplitudes with the same frequency unless the dynamics prohibits switching of the sign.
The average sign of the amplitude is zero and the chaotic system will not break the plus/minus symmetry.
On the other hand, a non-chaotic system can break this symmetry - an example is a particle which moves in a double well potential with friction.

On a level of a network composed of delayed coupled chaotic units with plus/minus symmetry, it has recently been shown \cite{Peleg} that many chaotic attractors can be found simultaneously with frozen sign of the amplitude of each chaotic unit. The number of such chaotic attractors scales exponentially with the size of the network.
However, for each isolated chaotic unit the sign of its amplitude is fixed by the initial condition, hence this network does not spontaneously break the plus/minus symmetry.

In this letter, we consider a network of nonlinear units with plus/minus symmetry.
Each isolated unit has chaotic trajectories which do not break this symmetry - the sign of the amplitudes of a chaotic trajectory is alternating.
We show, however, that in a network of such mutually coupled chaotic units the symmetry of each unit can be broken, spontaneously.
The sign of the amplitude of each unit is frozen while the entire network is still chaotic.
Furthermore, when the coupling is random or constructed from many patterns, there exists an exponentially large number of such chaotic states with frozen amplitude signs.

In the framework of statistical mechanics a randomly broken symmetry was found in spin-glass (SG) systems.
Each isolated spin is characterized by an alternating sign, stochastically driven by temperature.
A network of coupled spins with random coupling strengths, however, undergoes a phase transition, where the low temperature phase is characterized by a randomly frozen expectation value of each spin \cite{Nishimori,Binder,Mezard}.
The goal of this letter is to show that a similar phenomenon exists in a network where a discrete spin is replaced by a chaotic unit with alternating sign of its amplitude.

Results are exemplified with a network of delayed coupled modified Duffing oscillators \cite{Duffing 1} where the interactions are of the type of the pseudo-inverse model \cite{Pseudo-Inverse 1,Pseudo-Inverse 2}.
This chaotic network functions also as an associative memory with large basin of attractions surrounded by chaotic attractors.
The number of the pre-defined basins of attractions, the stored patterns, scale linearly with the size $N$ of the network whereas the total number of chaotic attractors increases exponentially with $N$.
Note that a network of chaotic maps with non-extensive logarithmic number of attractors was observed \cite{crisanti} for   uniform ferromagnetic interactions without delays.

\begin{figure}
\includegraphics[width=0.5\textwidth]{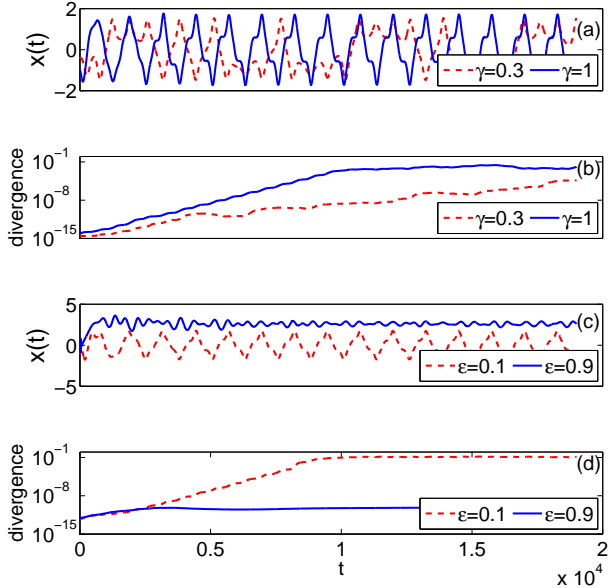}
\caption{
(a) A typical trajectory of a single MDE without self-feedback with
$\alpha=1, \beta=-1, \delta=0.2, \omega=1$.
(b) Divergence of two nearby initial conditions of a single MDE without self-feedback.
The divergence indicates chaotic behavior for both driving force amplitudes $\gamma=0.3$ and $\gamma=1$.
(c) Typical trajectories of a single MDE with self-feedback with $\alpha=1, \beta=-1, \gamma=1, \delta=0.2, \omega=1$. (d) Divergence of two nearby initial conditions as in (c).
In the weak coupling limit $(\epsilon=0.1)$ the trajectory is chaotic but not frozen, whereas in the strong coupling limit $(\epsilon=0.9)$ the trajectory is frozen but not chaotic. Although the stable points of the Duffing potential are at $x=\pm1$ the trajectory freezes at $x\approx 3$, indicating the freezing is not due to settling at the bottom of the potential well.
}
\label{fig1}
\end{figure}

The examined chaotic unit is a driven damped oscillator in a double well potential and is given explicitly by the Duffing equation
\begin{equation}\label{eq: regular Duffing equation}
\ddot x  = - \delta \dot x - \beta x - \alpha x^3  + \gamma \cos (\omega t)  \qquad (\alpha,\delta>0,\beta<0)
\end{equation}
which is known to be a simple chaotic model \cite{Duffing 1}.
The nature of the driving force in Eq. \eqref{eq: regular Duffing equation} breaks the inversion symmetry $(x\to-x)$ of the chaotic dynamics. In order to examine the possibility of broken symmetry on the network level where each chaotic unit is characterized by an alternating sign we introduce the following modified Duffing equation (MDE)
\begin{equation}\label{eq: modified Duffing equation}
\ddot x  = - \delta \dot x - \beta x - \alpha x^3  + \gamma \cos (\omega t) sgn(x) = \mathcal{F}(x)
\end{equation}
where $sgn$ is the signum function.
The dynamics of MDE is characterized by alternating sign amplitude, Fig. \ref{fig1}(a), and is also characterized by chaotic behavior, Fig. \ref{fig1}(b).
Specifically, the potential of the (modified) Duffing oscillator has stable points at $x=\pm\sqrt{\frac{|\beta|}{\alpha}}$ and an unstable point at $x=0$. When the driving force is too weak the oscillator will stay in one of the two wells while for strong driving force there are transitions from one well to the other.

Introducing a delayed self-feedback into the system
\begin{align}\label{eq: modified Duffing equation with feedback}
\ddot{x}(t) = (1-\epsilon)\mathcal{F}(x(t))+\epsilon x(t-\tau)
\end{align}
the trajectory may settle down after some transient time into a fixed sign amplitude.
There are mainly two dynamical regions characterized by the strength of the driving force. For a weak driving force, e.g. $\gamma=0.1$ in Fig. \ref{fig1}(c,d), the amplitude is  alternating and the trajectory is chaotic, while for a strong driving force, e.g. $\gamma=0.9$,  the sign of the amplitude is fixed and the dynamics is non-chaotic.
Quantitatively, the freezing of a trajectory is defined as
\begin{equation}
    \phi(x)=\frac{\langle x(t) \rangle^2 }{\langle x(t)^2 \rangle}
\end{equation}
where $\langle\cdot\rangle$ stands for time averaging.
For a uniform distribution with a given sign one can easily verify that $\phi=0.75$ and for a complete freezing $\phi=1$. Hence, freezing is characterized by
$\phi \in [0.75,1]$.

\begin{figure}
\includegraphics[width=0.5\textwidth]{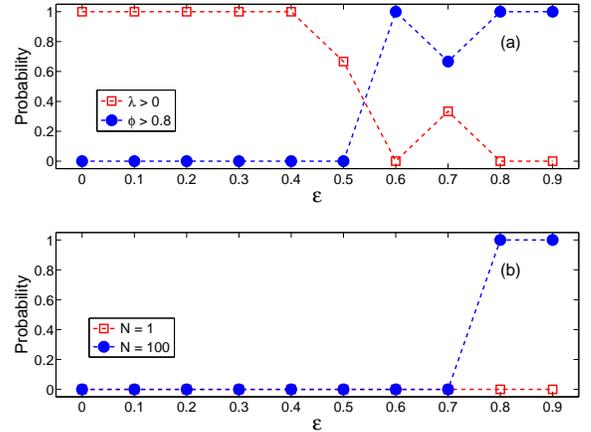}
\caption{
(a) Probability for a chaotic trajectory $(\lambda > 0)$ and the probability of freezing $(\phi > 0.8)$
as a function of $\epsilon$ for a single MDE with feedback, Eq. \eqref{eq: modified Duffing equation with feedback}, with
$\alpha=1, \beta=-1, \gamma=1, \delta=0.2, \omega=1$ and $\tau=0.2~$.
(b) Probability for the coexistence of chaotic trajectory and freezing as a function of $\epsilon$.
Parameters are: $\alpha=1, \beta=-1, \gamma=1, \delta=0.2, \omega=1$ and $\tau=0.2~$.
}
\label{fig2}
\end{figure}

The probability of finding a positive Lyapunov exponent $(\lambda>0)$ and the probability of freezing $(\phi > 0.8)$ in a MDE, Eq. \eqref{eq: modified Duffing equation with feedback}, as a function of the feedback strength $\epsilon$ are depicted in Fig. \ref{fig2}(a). In the weak coupling region each MDE, Eq. \eqref{eq: modified Duffing equation with feedback}, is chaotic and is not frozen while in the strong coupling region freezing emerges but chaotic dynamics disappear. There is a very limited intermediate region where both freezing and chaos can be found simultaneously with very low probability.

We introduce now a network of delayed coupled MDE units in which the above-mentioned postulate is violated. The fully connected network is defined by the interaction matrix $W$, which takes the form
\begin{equation}\label{eq: pseudo-inverse}
W_{ij}= \frac{1}{N}\sum_{\mu,\nu=1}^{P}\xi_{i}^{\mu}\left(C^{-1}\right)_{\mu \nu}\xi_{j}^{\nu} ~,~ C_{\mu\nu} =\frac{1}{N}\sum_{i=1}^{N}\xi_{i}^{\mu}\xi_{i}^{\nu}
\end{equation}
where $\xi^{\mu}_i= \pm 1$ with equal probability are the $P$ stored patterns ($\mu=1\ldots P, \ i=1\cdots N$) and $N$ is the size of the network.
In the prototypical model, units of the network are Ising spins $(S_i=\pm1)$ and the energy of the network is given by the Hamiltonian
\begin{equation}\label{eq: Hamiltonian}
H\ =\ -\frac{1}{2}\sum_{i \ne j}^N W_{ij} S_i S_j.
\end{equation}
The $P$ patterns are minima of the Hamiltonian, Eq. \eqref{eq: Hamiltonian}, and the pseudo-inverse network operates as an associative memory up to $P=N$ \cite{Pseudo-Inverse 2}, which is the upper bound of the capacity for Eq. \eqref{eq: Hamiltonian} with symmetric interactions. When the diagonal interactions $(W_{ii})$ are taken into account in Eq. \eqref{eq: Hamiltonian}, the network does not represent a Hamiltonian system and the capacity is reduced to $P=0.5N$ \cite{Pseudo-Inverse 1}. Nevertheless, the stability of the patterns is maximized in this case, since the matrix $W$ has $P$ eigenvalues equal to $1$ where the patterns are the corresponding eigenvectors, $W\bm{\xi}^{\mu} =\bm{\xi}^{\mu}$  (bold letters denote vectors), and $N-P$ zero eigenvalues.

The dynamics of MDEs, Eq. \eqref{eq: modified Duffing equation}, with the pseudo-inverse network, Eq. \eqref{eq: pseudo-inverse}, is given for unit $i$, for instance, by
\begin{equation}\label{eq: MDE network}
\ddot{x}_i(t) = (1-\epsilon)\mathcal{F}(x_i(t))+\epsilon \sum_{j=1}^{N} W_{ij} x_j(t-\tau)
\end{equation}
where the effective average self-coupling is $\langle\langle\epsilon W_{ii}\rangle\rangle = \epsilon \frac{P}{N}$ and $\langle\langle\cdot\rangle\rangle$ stands for an average over realizations of the interactions.
The freezing of a network is defined using an analog of the Edwards-Anderson order parameter which was introduced to measure the freezing in SG systems \cite{Nishimori,Mezard}
\begin{equation}\label{eq: Edwards-Anderson order parameter}
    q_{EA}=\sum_{i=1}^{N}\phi(x_i) \quad,\quad \phi(x_i)=\frac{\langle x_i(t) \rangle^2 }{\langle x_i(t)^2 \rangle}
\end{equation}
where $\langle\cdot\rangle$ stands for time averaging and $N$ is the size of the network.
Figure \ref{fig2}(b) examines the probability to find simultaneously chaotic dynamics $(\lambda>0)$ and freezing $(q_{EA}>0.8)$ as a function of the coupling strength, $\epsilon$. The probability is estimated using different initial conditions for a single MDE, Eq. \eqref{eq: modified Duffing equation with feedback}, and over different initial conditions and interactions for a network of size $N=100$, Eq. \eqref{eq: pseudo-inverse}. Results indicate that for a single MDE, independent of the coupling strength, both feature cannot be found simultaneously, in agreement with Fig. \ref{fig1}(c,d). However for a network a transition is observed where in the strong coupling region the dynamics is simultaneously chaotic and each MDE is frozen.

To examine the coexistence of freezing and chaotic dynamics we first turn to examine the dynamics in the vicinity of the patterns, $\bm{\xi}^{\mu}$, since we expect the pseudo-inverse network model to function as an associative memory.
In contrast to Hamiltonian systems, Eq. \eqref{eq: Hamiltonian}, the patterns are not fixed points of the chaotic dynamics, Eq. \eqref{eq: MDE network}, but fixed directions of the entire dynamical state of the network.
Setting the initial condition in the direction of one of the patterns
\begin{equation}
\bm{x}(t)=c(t)\bm{\xi}^\mu \qquad (c(t)>0,~0\le t\le\tau)
\end{equation}
and taking into account that $\xi_i^\mu=\pm1$ and $W\bm{\xi}^\mu = \bm{\xi}^\mu$,
one can find that the dynamics of the network is reduced to the projection of the state of the network on the given pattern only
\begin{align}\label{eq: i.c. one pattern}
&\ddot{c}(t)=(1-\epsilon)\mathcal{F}(c(t))+\epsilon c(t-\tau).
\end{align}
This equation describes merely the behavior of the coefficient $c(t)$ according to Eq. \eqref{eq: modified Duffing equation with feedback}. The overlap
\begin{equation}\label{eq: overlp}
Q^\mu(t)=\frac{\bm{x}(t)\cdot\bm{\xi}^\mu}{\|\bm{x}(t)\|\|\bm{\xi}^\mu\|}
\end{equation}
of the trajectory $\bm{x}(t)$ with the pattern $\bm{\xi}^\mu$ remains unity and the network maintains the same direction as its initial condition.

\begin{figure}
  \includegraphics[width=0.5\textwidth]{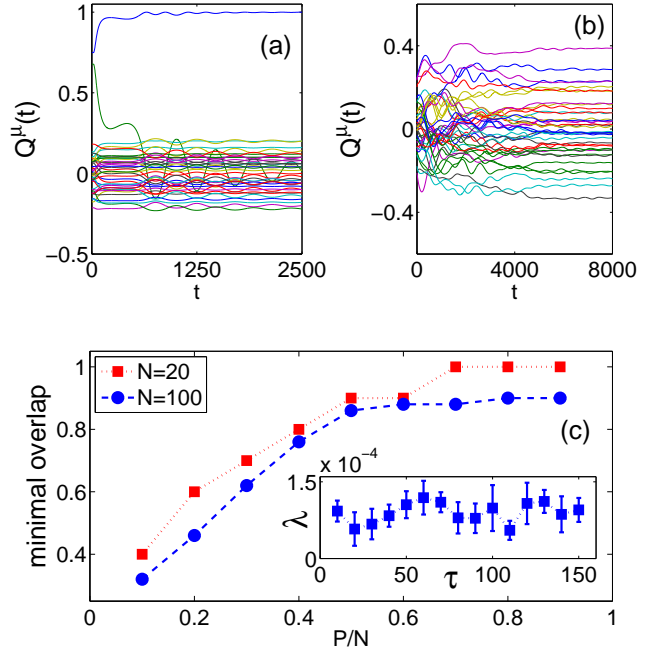}
   \caption{
  Freezing of the trajectory with respect to overlaps with the $P$ patterns with $N=50, P=0.4N$
  (a) for a Mattis trajectory
  (b) for a SG trajectory,
  where each point is averaged over window of $20\tau$ and the transient time is $\approx 5000$.
  (c) Basin of attraction of a Mattis state for $N=20,100$ as a function of $P$.
  Lines are guides to the eye only.
  Inset - Lyapunov Exponent inside the Mattis and SG attractors as a function of the delay time $\tau$.
  For all panels, parameters are: $\alpha=1, \beta=-1, \gamma=1, \delta=0.2, \omega=1$ and $\tau=0.2~$.
}
  \label{fig3}
\end{figure}

When the initial condition does not have a unit overlap with any of the patterns,
$|Q^{\mu}(t)|<1$, the state of the network can be expressed as
\begin{equation}\label{eq: i.c. many patterns}
\bm{x}(t)=\sum_{\mu=1}^{P} c_\mu(t)\bm{\xi}^\mu \quad (0\le t\le\tau).
\end{equation}
The dynamics in such a case cannot be written for each amplitude, $c_\mu(t)$, independently. The amplitudes are now coupled and as a consequence the overlaps $Q^\mu(t)$ are time dependent.
Nevertheless, the model is expected to function as an associative memory for the embedded patterns, $\bm{\xi}^{\mu}$, similar to other chaotic neural networks \cite{Airah1, Airah2}.
Specifically, initial conditions close to patterns $\mu$, for instance, are expected to converge closer to this pattern where $|Q^{\mu}|$ increases. Our predefined threshold to identify convergence to one pattern, known as a Mattis state \cite{Pseudo-Inverse 1,Pseudo-Inverse 2}, is $\left|\langle Q^\mu \rangle \right| > 0.9$ as depicted in Fig. \ref{fig3}(a).

Quantitatively, the basin of attraction is defined as the minimal initial overlap with a selected pattern $\mu$ (and random with the other patterns), which ensures asymptotically $\left|\langle Q^\mu \rangle \right| > 0.9$ with the same selected pattern.
Figure \ref{fig3}(c) indicates that the basin of attraction decreases as $P$ increases and it almost vanishes for $P/N>0.5$, since the initial overlap is $\left|\langle Q^\mu \rangle \right| > 0.9$.
The basin of attraction also increases with the system size $N$, indicating that asymptotically for $P/N<0.5$ it is a macroscopic quantity.

\begin{figure}
\includegraphics[width=0.5\textwidth]{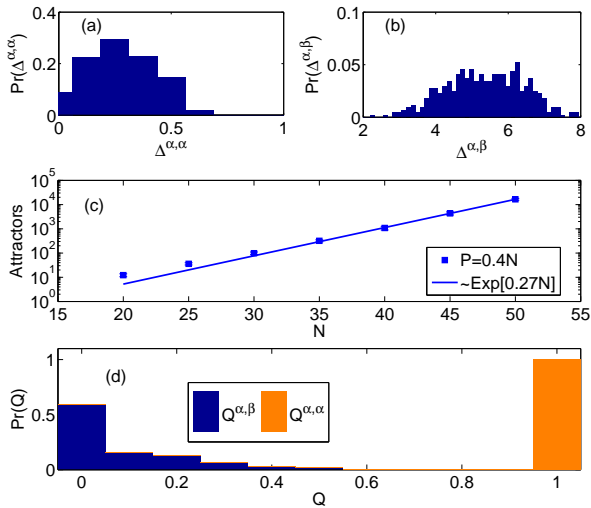}
\caption{Histograms of the Manhattan distances with $N=40,~P=0.4N$
{(a)} between different sections along a trajectory $\Delta^{\alpha,\alpha}$
{(b)} between different trajectories $\Delta^{\alpha,\beta}$.
Results indicate a substantial gap between the maximal value of $\Delta^{\alpha,\alpha}$ and the minimal value of $\Delta^{\alpha,\beta}$.
{(c)}
Number of SG attractors scales as $\approx0.02\exp(0.27N)$ for $P=0.4N$. Error bars are smaller than size of symbols.
{(d)}
Relative freezing between different sections along a trajectory $Q^{\alpha,\alpha}$ and between different trajectories $Q^{\alpha,\beta}$ for $N=40,P=0.4N$ and $\epsilon=0.9$.
For all panels, parameters are: $\alpha=1, \beta=-1, \gamma=1, \delta=0.2, \omega=1$ and $\tau=0.2~$.
}
\label{fig4}
\end{figure}

For large enough $N$, random initial conditions, characterized by an overlap $\sim 1/\sqrt{N}$ with all patterns, are typically out of the basin of attraction of a Mattis state.
The main  puzzle is whether the trajectory is frozen, $\left( q_{EA}>0.8 \right)$, when there is no dominated overlap with any of the patterns.
Figure \ref{fig3}(b) indicates that this is indeed the case, where after a transient time overlaps with all the patterns are frozen and their values are small, $\left|\langle Q^{\mu} \rangle\right|<3/\sqrt{N}$. These attractors are identified as SG attractors characterized by the coexistence of random freezing of the network $\left(q_{EA}>0.8\right)$ and
$\left|\langle Q^\mu \rangle \right| \lesssim 3/\sqrt{N}$. However, the trajectory is still chaotic and the Lyapunov exponent is on the average the same as for the Mattis states, inset of Fig. \ref{fig3}(c).

To count the number of SG chaotic attractors, we use the same procedure as in \cite{Peleg}.
After a transient time the average overlaps are calculated and the pattern with largest overlap is labeled $P$, the pattern with the second largest overlap as $P-1$ $\ldots$ and the pattern with smallest overlap as $1$.
The vector $\bm{R}$ of the $P$ patterns with their ranking, is a permutation of $(1,2,\ldots,P)$.
We now define the Manhattan distance \cite{Distance} between two ranking vectors by
\begin{equation}\label{eq: Manhattan Distance}
    \Delta^{\alpha,\beta}=\frac{1}{P}\sum_{\mu=1}^{P} \left|R_\mu^\alpha-R_\mu^\beta\right|~.
\end{equation}
One can easily verify that $0\le\Delta^{\alpha,\beta}\le P/2$ and when correlations between two random ranking vectors are neglected one can find that the average Manhattan distance is $\frac{1}{3}\left(P-\frac{1}{P}\right)$.
However, simulation results, Fig. \ref{fig4}(b), indicate a deviation from this expected value.
The distribution of the Manhattan distances obtained between different segments along one trajectory, $\Delta^{\alpha,\alpha}$, Fig. \ref{fig4}(a), is much less than $2$, indicating a substantial freezing.
In contrast, the Manhattan distance between ranking vectors of two different frozen trajectories, after a long transient time is much larger than $2$, $\Delta^{\alpha,\beta}\approx 5.3$ as expected for $P=16$.
This gap is the crucial quantity which enables us to approximate the number of SG chaotic attractors, as it represents a quantitative measure to distinguish between different chaotic frozen SG attractors.
Specifically, we first build an initial ensemble of $N_0$ different ranking vectors where the Manhattan distance  between any pair $\Delta^{\alpha,\beta}>2$.
Next, starting from $N_1$ new random initial conditions we find $N_1$ new frozen trajectories.
We calculate the fraction $\rho$ of the new frozen trajectories which has already appeared in the initial ensemble, i.e. their Manhattan distance to one of the $N_0$ trajectories is less than $2$.
The number of attractors is then approximated as
$$\frac{N_0}{\rho}.$$
This estimation was done under the assumption that the basin of attraction of all chaotic attractors is similar, as was confirmed in simulations where starting from large random initial conditions the probability to fall into each attractor was similar.
Using the above method we found that the number of chaotic attractors obeys the scaling $\sim\exp(0.27N)$ for $\gamma=1$ as depicted in Fig. \ref{fig4}(c).

In analogy to SG systems, the relative direction between two $N$ dimensional trajectories $(\alpha$ and $\beta)$ is defined similarly to Eq. \eqref{eq: overlp} by
\begin{equation}\label{}
    Q^{\alpha,\beta}(t)=
    \frac{\bm{x}^\alpha(t)\cdot\bm{x}^\beta(t)}{\|\bm{x}^\alpha(t)\|\|\bm{x}^\beta(t)\|}
\end{equation}
where $|Q^{\alpha,\beta}| \le 1$.
Figure \ref{fig4}(d) indicates that the overlap $Q^{\alpha,\alpha}$ between different segments along a trajectory is above $0.9$ indicating substantial freezing, while the overlap $Q^{\alpha,\beta}$ between different trajectories remains small as different trajectories are pointing in a random direction in the $N$ dimensional space. (For $N=40$ in Fig. \ref{fig4}(d) the expected overlap is $1/\sqrt{40} \approx 0.16~$.)

In conclusion, the emergence of symmetry breaking in the form of random freezing of the chaotic attractors is a consequence of cooperation among the entire network units.
In contrast, single unit does not break the symmetry due to its chaotic motion.
We present results for a specific set of parameters of the MDE, Eq. \eqref{eq: modified Duffing equation with feedback}, however, similar qualitative behavior have been found also for different driving amplitudes such as $\gamma=0.3$ and $\gamma=0.5$ and various time delays. This phenomenon of breaking symmetry is not limited to pseudo-inverse types of interactions, Eq. \eqref{eq: pseudo-inverse}, and was observed in our simulations also for random interaction matrices.
However, the scaling of the number of attractors in such networks deserves further research, as well as the the necessary conditions on the types of the interactions leading to the coexistence of symmetry breaking and random freezing.

\end{document}